\documentclass[twocolumn]{aastex62}
\journalinfo{\@Draft}
\usepackage{natbib}
\usepackage{xcolor}
\usepackage{color}
\usepackage{comment}
\usepackage{amsmath}
\definecolor{color}{RGB}{25,25,112}
\definecolor{negro}{RGB}{0,0,0}
\definecolor{colorurl}{RGB}{25,25,112}
\definecolor{royalfuchsia}{rgb}{0.79, 0.17, 0.57}
\definecolor{red}{rgb}{1.0, 0., 0.}
\renewcommand\tablename{Table}

\hypersetup{
  colorlinks,
  citecolor=color,
  linkcolor=color,
  urlcolor=colorurl}
\shorttitle{BROADENING CORRELATION TO OPTICAL BY OBSERVATIONS OF THE GRB 180325A}

\begin{document}

\title{
MODELLING THE PROMPT OPTICAL EMISSION OF GRB 180325A: THE EVOLUTION OF A SPIKE FROM THE OPTICAL TO GAMMA-RAYS\\
}

\author{Becerra,~R.~L.}
\affiliation{Instituto de Ciencias Nucleares,  Universidad Nacional Aut\'onoma de M\'exico, Apartado Postal 70-543, 04510 CDMX, M\'exico}

\author{De Colle,~F.}
\affiliation{Instituto de Ciencias Nucleares,  Universidad Nacional Aut\'onoma de M\'exico, Apartado Postal 70-543, 04510 CDMX, M\'exico}

\author{Cant\'o,~J.}
\affiliation{Instituto de Astronom{\'\i}a, Universidad Nacional Aut\'onoma de M\'exico, Apartado Postal 70-264, 04510 CDMX, M\'exico}

\author{Lizano,~S.}
\affiliation{Instituto de Radioastronom{\'\i}a y Astrof{\'\i}sica, Universidad Nacional Aut\'onoma de M\'exico, 58089 Morelia, M\'exico}

\author{Gonz\'alez,~R.~F.}
\affiliation{Instituto de Radioastronom{\'\i}a y Astrof{\'\i}sica, Universidad Nacional Aut\'onoma de M\'exico, 58089 Morelia, M\'exico}

\author{Granot, J.}
\affiliation{Department of Natural Sciences, The Open University of Israel, P.O. Box 808, Ra'anana 43537, Israel ; Department of Physics, The George Washington University, Washington, DC 20052, USA}

\author{Klotz,~A.}
\affiliation{IRAP, Universit\'e de Toulouse, CNRS, CNES, UPS, (Toulouse), France}

\author{Watson,~A.~M.}
\affiliation{Instituto de Astronom{\'\i}a, Universidad Nacional Aut\'onoma de M\'exico, Apartado Postal 70-264, 04510 CDMX, M\'exico}

\author{Fraija,~N.}
\affiliation{Instituto de Astronom{\'\i}a, Universidad Nacional Aut\'onoma de M\'exico, Apartado Postal 70-264, 04510 CDMX, M\'exico}

\author{Araudo, A.~T.}
\affiliation{Astronomical Institute of the Czech Academy of Sciences, Bocni II 1401, Prague, CZ-14100 Czech Republic}
\affiliation{ELI Beamlines, Institute of Physics, Czech Academy of Sciences,
25241 Doln{\'\i} B\v re\v zany, Czech Republic}

\author{Troja,~E.}
\affiliation{Astrophysics Science Division, NASA Goddard Space Flight Center, 8800 Greenbelt Road, Greenbelt, MD 20771, USA}
\affiliation{Department of Astronomy, University of Maryland, College Park, MD 20742-4111, USA}

\author{Atteia,~J.~L.}
\affiliation{IRAP, Universit\'e de Toulouse, CNRS, CNES, UPS, (Toulouse), France}

\author{Lee,~W.~H.}
\affiliation{Instituto de Astronom{\'\i}a, Universidad Nacional Aut\'onoma de M\'exico, Apartado Postal 70-264, 04510 CDMX, M\'exico}

\author{Turpin,~D.}
\affiliation{Universit\'e Paris-Saclay, CNRS, CEA, D\'epartement d'Astrophysique, Astrophysique, Instrumentation et Mod\'elisation de Paris-Saclay, 91191, Gif-sur-Yvette, France.}

\author{Bloom,~J.~S.}
\affiliation{Department of Astronomy, University of California, Berkeley, CA, 94720-3411, USA}

\author{Boer,~M.}
\affiliation{ARTEMIS, UMR 7250 (CNRS/OCA/UNS), boulevard de l'Observatoire, BP 4229, F 06304 Nice Cedex, France}

\author{Butler,~N.~R.}
\affiliation{ School of Earth and Space Exploration, Arizona State University, Tempe, AZ 85287, USA}

\author{Gonz\'alez,~J.~J.}
\affiliation{Instituto de Astronom{\'\i}a, Universidad Nacional Aut\'onoma de M\'exico, Apartado Postal 70-264, 04510 CDMX, M\'exico}

\author{Kutyrev,~A.~S.}
\affiliation{Astrophysics Science Division, NASA Goddard Space Flight Center, 8800 Greenbelt Road, Greenbelt, MD 20771, USA}
\affiliation{Department of Astronomy, University of Maryland, College Park, MD 20742-4111, USA}

\author{Prochaska,~J.~X.}
\affiliation{Department of Astronomy and Astrophysics, UCO/Lick Observatory, University of California, 1156 High Street, Santa Cruz, CA 95064, USA}

\author{Ramirez-Ruiz,~E.}
\affiliation{Department of Astronomy and Astrophysics, UCO/Lick Observatory, University of California, 1156 High Street, Santa Cruz, CA 95064, USA}

\author{Richer,~M.~G.}
\affiliation{Instituto de Astronom{\'\i}a, Universidad Nacional Aut\'onoma de M\'exico, Unidad Acad\'emica en Ensenada, 22860 Ensenada, BC, Mexico}

\author{Rom\'an-Z\'u\~niga,~C.~G.}
\affiliation{Instituto de Astronom{\'\i}a, Universidad Nacional Aut\'onoma de M\'exico, Unidad Acad\'emica en Ensenada, 22860 Ensenada, BC, Mexico}

\begin{abstract}

The transition from prompt to the afterglow emission is one of the most exciting and least understood  phases in gamma-ray bursts (GRBs).  Correlations among optical, X-ray and gamma-ray emission in GRBs have been explored, to attempt to answer whether the earliest optical emission
comes from internal and/or external shocks.  We present optical photometric observations of GRB 180325A collected with the TAROT and RATIR ground-based telescopes. These observations show two strong optical flashes with separate peaks at $\sim50\;$s and $\sim120\;$s, followed by a temporally extended optical emission. We also present X-rays and gamma-ray observations of GRB 180325A, detected by the Burst Alert Telescope (BAT) and X-ray Telescope (XRT), on the Neil Gehrels Swift observatory, which both observed a narrow flash at $\sim80\;$s. We show that the prompt gamma- and X-ray early emission shares similar temporal and spectral features consistent with internal dissipation within the relativistic outflow (e.g. by internal shocks or magnetic reconnection), while the early optical flashes are likely generated by the reverse shock that decelerates the ejecta as it sweeps up the external medium.  

\end{abstract}
\begin{center}
\end{center}

\section{Introduction}
\label{sec:introduction}

Gamma-ray bursts (GRBs) are the most luminous events in the Universe. GRBs are classified according to their duration: short GRBs have a duration $T_{90}\lesssim 2$~s (where $T_{90}$ is defined as the time over which 90\% of the gamma-ray photons are detected) and are associated with the coalescence of compact objects \citep[two neutron stars or a neutron star/black hole binary system;][]{1986ApJ...308L..43P,1989Natur.340..126E,2017ApJ...848L..13A}, while long GRBs are associated with the collapse of massive stars and the formation of a black hole or a magnetar, and have a typical duration $\gtrsim 2$~s \citep{1993ApJ...405..273W,1999ApJ...524..262M,2003Natur.423..847H}.

One of main features of the prompt GRB phase is its large variability. In the ``standard fireball model'' of GRBs, the central engine powers a jet with variable velocity. This leads to the formation of internal shocks in which collisions give rise to the observed rapid variability \citep[e.g.,][]{1997ApJ...490...92K}. An alternative dissipation mechanism is magnetic reconnection in a Poynting-flux dominated outflow. In addition to this 
short timescale variability, about 30\% of GRBs show X-rays flares \citep[e.g.,][]{chincarini07, falcone07,2005SSRv..120..165B} which are not associated with bright high-energy (MeV to GeV) emission \citep{2015ApJ...803...10T,2020arXiv200610291F}. 
These X-ray flares share several temporal and spectral properties with the gamma-ray pulses observed during the prompt emission \citep[e.g.][]{Krimm07,Margutti11}, 
suggesting that they may be produced by late engine activity. For instance, both gamma-ray and X-ray sharp peaks are asymmetric, with fast rise and a slower decay. They both follow a similar hard-to-soft evolution, have spectra typically well fitted by a Band function and present a similar spectral lag  \citep{chincarini10}. In addition, the observed small values of $\Delta t/t_p \approx 0.1$ (in which $\Delta t$ is the duration of the pulse and $t_p$ the time of the peak, see, e.g., \citealt{2005ApJ...631..429I}) and the presence of a continuum with a similar slope before and after the X-ray flare, while distinguishing them from the prompt GRB emission, imply that they are not typically associated with the external shock that is produced when the outflow is decelerated by the surrounding medium. An alternative scenario is that the X-ray flare originate in late-time sporadic magnetic reconnection events \citep{Giannios06}.

According to the fireball model, the external shock is when the outflow interacts with the circumstellar medium forming a forward and a reverse shock \citep{1993ApJ...405..278M,SP95}. While the long-lasting ``afterglow" emission in wavelengths ranging usually from radio bands to X-rays is attributed to synchrotron radiation from the
forward shock, the main observational signature of the reverse shock is typically a strong optical flash observable in the very early stages of the afterglow \citep[e.g.,][]{SP99,2000ApJ...545..807K,2015ApJ...804..105F}. This flash is due to synchrotron emission from the shocked ejecta and is typically expected to produce a single peak in the light curve \citep[e.g., see][]{2016ApJ...818..190F,2018ApJ...859...70F,2019ApJ...881...12B}. Bright emission requires a moderately magnetised ejecta. Optical flashes or flares are much less common  \citep[e.g.,][]{kruhler09, li12} and typically not correlated with X-ray sharp peaks \citep[e.g.,][]{2015ApJ...803...10T}.

In this paper, we present optical observations of GRB 180325A with the TAROT and RATIR ground-based robotic telescopes. These observations show strong optical flashes with two distinct peaks at $\sim50\;$s and $\sim120\;$s, followed by temporally extended optical emission.

In order to analyse our optical data, we present other optical observations, as well as X-ray and gamma-ray observations with the Burst Alert Telescope (BAT) and X-Ray Telescope (XRT), on the Neil Gehrels Swift Observatory \cite{2004ApJ...611.1005G}. The gamma-ray and X-ray emission show narrow peaks at $\sim80\;$s. Using a semi-analytic model we show that the gamma- and X-ray early emission that shares temporal and spectral features is consistent with production by internal shocks, while the the early optical flash is likely generated by the reverse shock. We also discuss the possibility the the multi-wavelength peaks have a common origin. We show that this possibility is very unlikely for both  observational and theoretical reasons, although it cannot be completely excluded. 

This paper is organised as follows. In \S\ref{sec:observations}, we present the observations made by {\itshape Swift}, TAROT, RATIR, and other telescopes. In \S\ref{sec:analysis} we present a temporal and spectral analysis of the data. In \S\ref{sec:spike} we outline
different scenarios to try to explain the observed spikes, using a semi-analytic model for internal shocks. In \S\ref{sec:discussion} we summarise our results and discuss their implications.

\section{Observations}
\label{sec:observations}
\subsection{Neil Gehrels Swift Observatory}
\label{sec:swift}

The {\itshape Swift}/BAT instrument \citep{2005SSRv..120..143B} triggered on GRB 180225A at $T =$ 2018 March 25 01:53:02.84 UTC (trigger 817564). \citet{22532} reported a total duration $T_{90}=120$~s. {\itshape Swift}/BAT observed an initial FRED (fast rise and exponential decay) pulse from $T+0$ to $T+10$~s and then a stronger second pulse starting at about $T+70$~s, peaking at about $T+80$~s, and lasting until about $T+110$~s \citep{22545}. 
The 15--150 keV fluence was $(6.5\pm0.2) \times 10^{-6}~\mathrm{erg\,cm^{-2}}$ and the 15--350 keV duration $T_{90}$ was $94.1 \pm 1.5$~s \citep{22545}, clearly identifying GRB 180325A as a long GRB.

The {\itshape Swift}/XRT instrument \citep{2005SSRv..120..165B} started observing the field at 2018 March 25 01:54:16.2 UTC ($T+73.4$~s) and detected a bright, fading source at 10:29:42.56 +24:27:49.0 J2000 with a 90\% uncertainty radius of 1.5 arcsec \citep{22532,22539}.

The {\itshape Swift}/UVOT began settled observations of the field of GRB 180325A at $T + 82$~s and detected a fading source consistent with the enhanced XRT position \citep{2018GCN.22549....1M}.

\subsection{TAROT Observations}
\label{sec:observationstarot}

The TAROT Calern telescope \citep{2008PASP..120.1298K} received the BAT GRB position 
GCN alert packet at 2018 March 25 01:53:17.52 UTC ($T+14.68$~s). It immediately slewed to the BAT position, and began its first exposure at 01:53:28.94 UTC ($T + 26.10$~s). The observations were taken in the $C$ filter which is fairly close to $r$ for sources with neutral colours.

The first TAROT image was exposed from $T+26$ to $T+86$~s with the tracking speed adjusted to obtain a small trail eleven pixels long. This technique is used to obtain
continuous temporal information during the exposure \cite[see, e.g.,][]{2006A&A...451L..39K}. The spatial sampling was 3.29 arcsec/pixel and the FWHM of stars (perpendicular to the trail) was 2.8 pixels. Subsequent images were taken with standard sidereal tracking.

To calibrate the photometry we use NOMAD-1 1144-0181764 as reference star and adopted $r=14.53$ and $r-i=+0.41$ from SDSS DR9. This star was chosen because its color is very close to the color index $r-i=+0.39$ of the optical counterpart measured about one hour after the trigger by \cite{2018GCN.22544....1S} and \cite{22537}. The $r$ magnitude for the reference star was converted into flux density $F_{\rm ref} = 5.60$~mJy. The flux ratio between the afterglow and the reference star was determined by subtracting a scaled and shifted subframe around the reference star from a subframe around the optical counterpart and minimising the RMS residual.

Table~\ref{tab:datostarot} gives TAROT photometry. For each exposure, it gives the initial time $t_\mathrm{i}$ and the final time $t_\mathrm{f}$ (relative to $T$), the AB $r$ magnitude (not corrected for Galactic extinction), and the 1$\sigma$ total uncertainties (including both statistical and systematic contributions).
For TAROT, the exposure time is simply $t_\mathrm{f}-t_\mathrm{i}$. 

The TAROT data show non-detections from 26 to 42 seconds, followed by detections of a flash from 42 to 53 seconds, followed again by non-detections from 53 to 69 seconds. After 69 seconds, TAROT detects a rapidly-rising flash with then fades more slowly.

The first optical flash is not seen in the BAT light curve and unfortunately occurs before the start of XRT observations. The second optical flash is nearly simultanous with flashes seen in both the BAT and XRT light curves.

\begin{deluxetable}{ccr}
\tablecaption{TAROT Observations of GRB 180325A\label{tab:datostarot}}
\tablehead{\colhead{$t_i$ ($s$)}& \colhead{$t_f$ ($s$)}&\colhead{$r$ (AB)}}
\startdata
25.92 & 31.32 & $> 18.69 $\\
31.32 & 37.08 & $> 18.69 $\\
37.08 & 42.48 & $> 18.69 $\\
42.48 & 47.88 & $ 17.80 \pm 0.38 $\\
47.88 & 53.28 & $ 17.36 \pm 0.31 $\\
53.28 & 58.68 & $> 18.69 $\\
58.68 & 64.08 & $> 18.69 $\\
64.08 & 69.48 & $> 18.69 $\\
69.48 & 75.24 & $ 18.55 \pm 0.43 $\\
75.24 & 80.64 & $ 17.65 \pm 0.27 $\\
80.64 & 86.04 & $ 17.13 \pm 0.25 $\\
99.72 & 129.96 & $ 16.34 \pm 0.09 $\\
140.76 & 170.64 & $ 17.32 \pm 0.19 $\\
181.44 & 293.40 & $ 17.78 \pm 0.20 $\\
303.48 & 393.48 & $ 18.16 \pm 0.31 $\\
742.32 & 1033.92 & $ 18.74 \pm 0.31 $\\
1062.36 & 1571.04 & $ 18.92 \pm 0.36 $\\
1581.12 & 4263.48 & $ 19.14 \pm 0.22 $\\
4464.00 & 6028.56 & $ 19.42 \pm 0.35 $\\

\enddata
\end{deluxetable}

\subsection{RATIR Observations}
\label{sec:observationsratir}

RATIR is a four-channel simultaneous optical and near-infrared imager mounted on the 1.5 meter Harold L. Johnson Telescope at the Observatorio Astron\'omico Nacional on Sierra San Pedro M\'artir in Baja California, Mexico. RATIR responds autonomously to GRB triggers from the Swift satellite and obtains simultaneous photometry in $riZJ$ or $riYH$ \citep{2012SPIE.8446E..10B,2012SPIE.8444E..5LW,2015MNRAS.449.2919L}. Unfortunately, the $ZYJH$ detectors were not in service during our observations. Therefore, we only report the observations carried on by $r$ and $i$ filters.

GRB 180325A occurred just before local sunset at the Observatorio Astron\'omico Nacional, and our observations with RATIR did not start until the end of nautical twilight \citep{22537}. On the first night of 2018 March 25 we observed from 02:54 UTC to 10:55 UTC ($T + 1.02$ to $T+9.03$ hours) and obtained 288 pairs of simultaneous exposures each of 80~s in $r$ and $i$. On the second night of 2018 March 26 we observed from 03:02 UTC to 10:47 UTC ($T + 25.17$ to $T+32.90$ hours) and obtained 272 pairs of simultaneous exposures each of 80~s in $r$ and $i$. 

Our reduction pipeline performs bias subtraction and flat-field correction, followed by astrometric calibration using the astrometry.net software \citep{2010AJ....139.1782L}, iterative sky-subtraction, coaddition using SWARP \citep{2002ASPC..281..228B}, and source detection using SEXTRACTOR \citep{1996A&AS..117..393B}. We calibrate against SDSS. The systematic calibration error is about 1\%.

\begin{figure*}
\centering
 \includegraphics[width=0.8\textwidth]{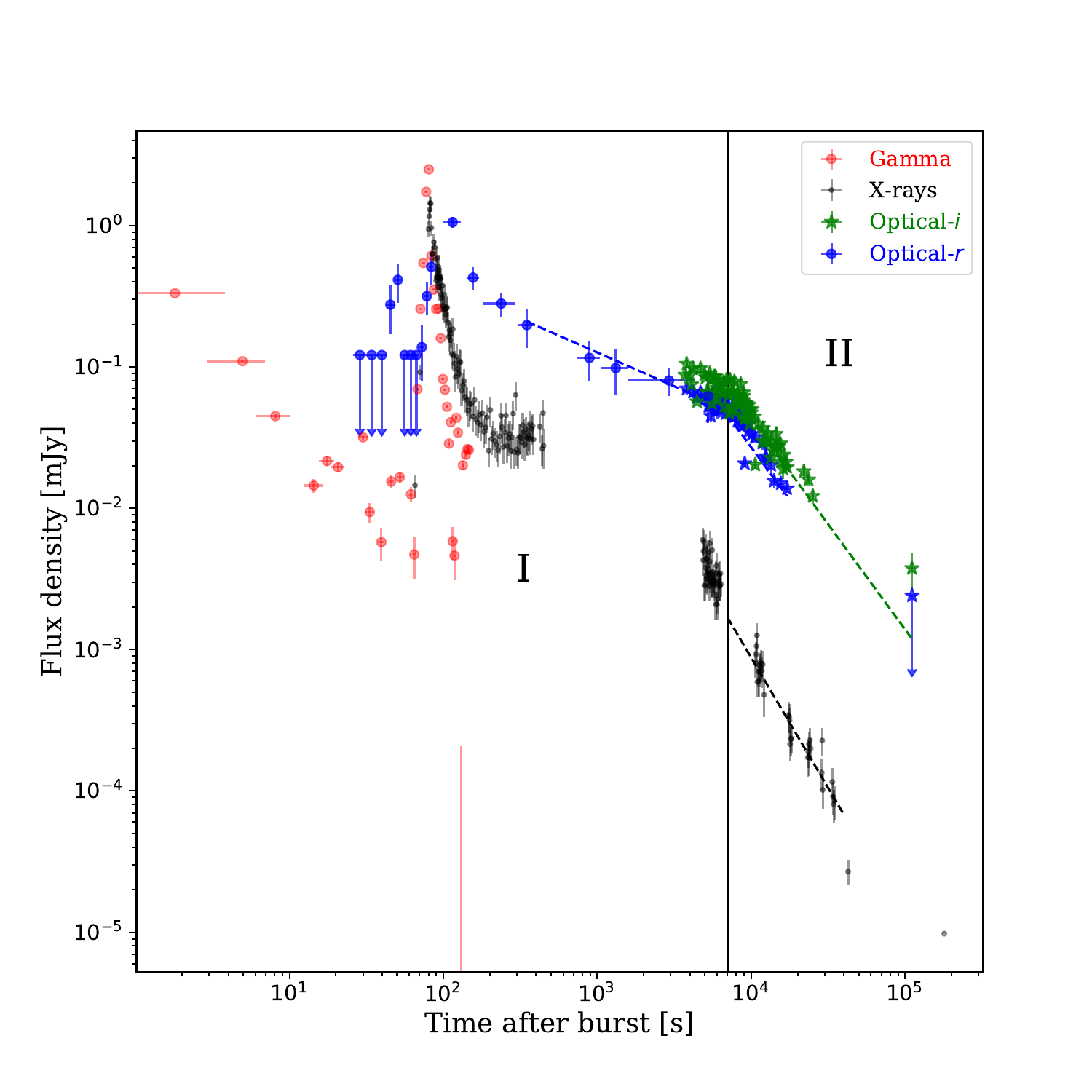}
 \caption{Light curves and broken power-law fits of GRB 180325A in $r$ from TAROT (blue points), in $r$ and $i$ from RATIR (blue and green stars), X-rays from {\itshape Swift}/XRT (black points) at 1~keV and gamma-rays from {\itshape Swift}/BAT (red points). We divide the light curve of GRB 180325A in different epochs. Epoch I ($t<7000$ s) is related with late central activity for the spike and the subsequent plateau. On the other hand, Epoch II ($t>700$0 s) is a typical afterglow decay.}
 \label{fig:observations}
\end{figure*}

Table~\ref{tab:datosratir} gives our RATIR photometry. For each image it gives the initial time $t_i$, the final time $t_f$, the total exposure time $t_{\rm exp}$, the $r$ and $i$ AB magnitudes (not corrected for Galactic extinction), and their 1$\sigma$ total uncertainties (including both statistical and systematic contributions). Figure~\ref{fig:observations} shows the light curve of GRB 180325A from TAROT and RATIR.

\begin{deluxetable}{rrrcc}
\tablewidth{\columnwidth}
\tablecaption{RATIR observations of GRB 180325A\label{tab:datosratir}}
\tablehead{\colhead{$t_i$ (s)}& \colhead{$t_f$ (s)}& \colhead{$t_{\rm exp}$ (s)}&\colhead{$r$ (AB)}&\colhead{$i$ (AB)}}
\startdata
3674.8 & 5203.1 & 720 & 19.44 $\pm$ 0.04 & 19.51 $\pm$ 0.03 \\
5235.8 & 6751.3 & 1120 & 19.58 $\pm$ 0.02 & 19.27 $\pm$ 0.02 \\
6781.1 & 8290.6 & 960 & 19.71 $\pm$ 0.02 & 19.35 $\pm$ 0.02 \\
8326.8 & 9823.9 & 1120 & 20.61 $\pm$ 0.05 & 19.54 $\pm$ 0.02 \\
9858.2 & 11269.4 & 400 & 20.15 $\pm$ 0.09 & 20.63 $\pm$ 0.07 \\
11808.3 & 13222.8 & 800 & 20.51 $\pm$ 0.05 & \nodata \\
11808.4 & 13222.8 & 800 & \nodata & 20.22 $\pm$ 0.05 \\
13351.6 & 14739.4 & 640 & 20.92 $\pm$ 0.11 & 20.40 $\pm$ 0.06 \\
14869.0 & 16270.9 & 720 & 20.96 $\pm$ 0.07 & \nodata \\
14870.0 & 16270.9 & 720 & \nodata & 20.57 $\pm$ 0.06 \\
16399.7 & 17790.6 & 1120 & 21.05 $\pm$ 0.07 & 20.67 $\pm$ 0.05 \\
21306.4 & 22599.7 & 400 & \nodata & 20.75 $\pm$ 0.12 \\
22812.1 & 24326.0 & 1040 & \nodata & 20.90 $\pm$ 0.09 \\
24351.6 & 25832.6 & 880 & \nodata & 21.18 $\pm$ 0.11 \\
103125.3 & 118366.9 & 11680 & \nodata & 22.46 $\pm$ 0.29 \\
103449.8 & 118366.8 & 11680 & $>$22.94 & \nodata \\

\enddata
\end{deluxetable}

\subsection{Other Observations}
\label{sec:observationsterrestrial}

\cite{2018GCN.22535....1H} obtained spectroscopy with NOT/ALFOSC. They identified several absorption features, including Mg II, Fe II, Al II, C IV, Si II and determined a common redshift of $z=2.25$ for them. 
\cite{2018GCN.22555....1D} obtained spectroscopy with VLT/X-Shooter. They identified absorption and emission lines at $z=2.248$ and confirmed the presence of a strong double intervening system at $z = 2.041/2.043$. 

\cite{2018ApJ...860L..21Z}  determined that the host galaxy is from the main-sequence of star-forming galaxies using observations at four different epochs with NOT, VLT/X-shooter, and GROND. 

\cite{2018GCN.22546....1F} estimated the following rest-frame parameters using Konus-Wind: the isotropic \textcolor{red}{photon} energy release $E_{\rm iso}$ is  about $2.3\times10^{53}$ erg, the peak luminosity $L_{\rm iso}$ is about $3.2\times10^{53}$ erg/s, and the rest-frame peak energy of the time-integrated spectrum, $E_{p,z}$, is ~995 keV.

Further optical observations were reported by \cite{2018GCN.22551....1M,2018GCN.22541....1S,2018GCN.22544....1S}.

\section{Spectral and Temporal Analysis}
\label{sec:analysis}


\subsection{Spectral Analysis}
\label{sec:spectrum}

\begin{figure}
\centering
 \includegraphics[width=0.45\textwidth]{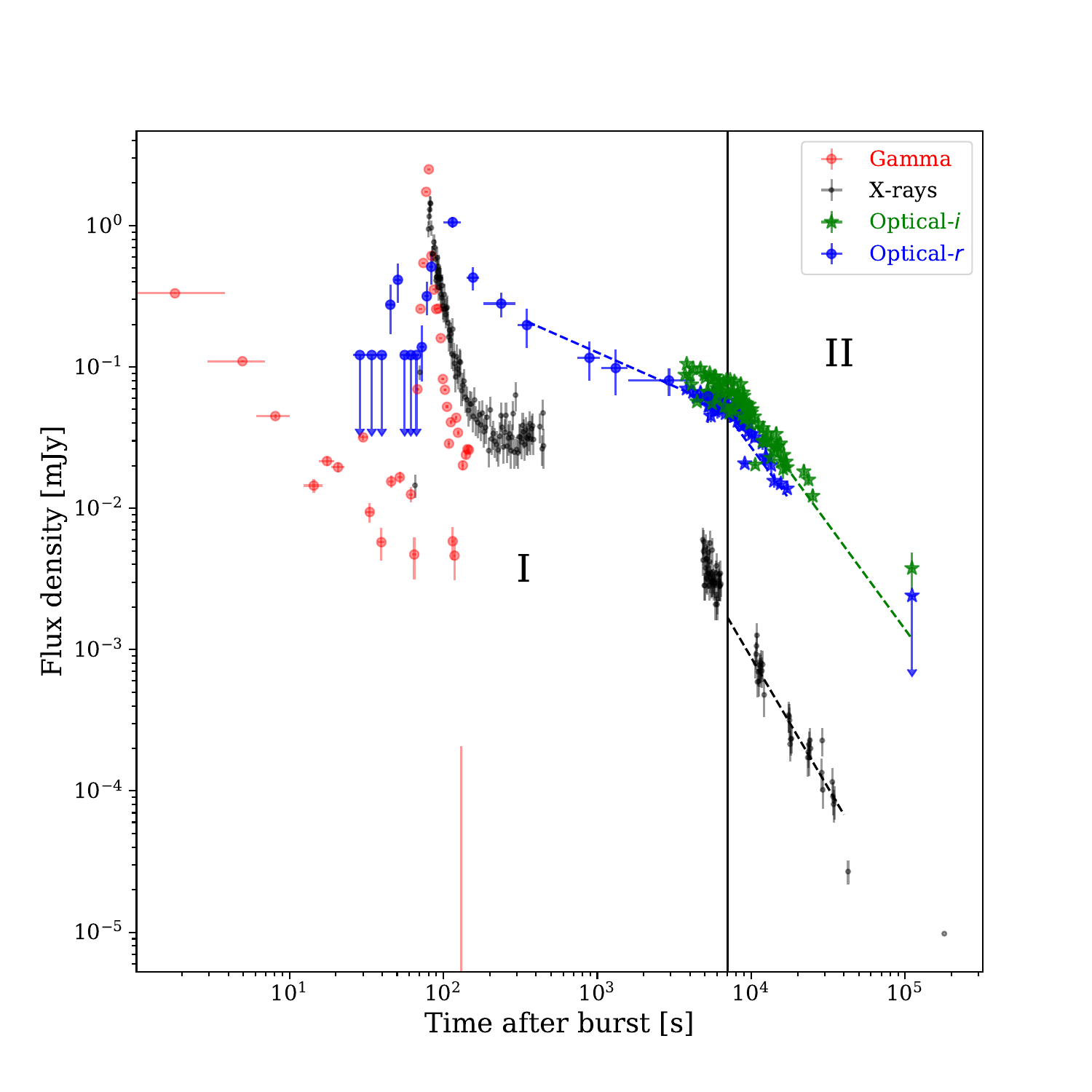}
 \caption{The SED of GRB 180620A at $T+10000$~s from X-rays to the optical. The data are from RATIR, {\itshape Swift}/UVOT, and {\itshape Swift}/XRT. The line is a Xspec model that takes into account the effects of reddening and absorption by dust.}
 \label{fig:sed}
\end{figure}

We constructed the spectral energy distribution (SED) at $T+10000$~s (just after the start of the afterglow phase) by
retrieving the {\itshape Swift}/XRT X-ray from the online repository \footnote{http://www.swift.ac.uk/xrt\_spectra/} \citep{2009MNRAS.397.1177E}. We added optical observations in the $r$ and $i$ filters of RATIR at $T+10000$~s. 
Figure~\ref{fig:sed} shows the resulting SED.

We fitted the SED with a spectral power-law $F_\nu\propto \nu^{-\beta}$, in which $F_\nu$ is the flux density, $\nu$ is the frequency, and $\beta$ is the spectral index.
From the X-ray to the optical, the SED can be fitted with a simple power law with a spectral index of \textbf{$\beta=0.45\pm0.01$}. Under our assumption of a thin-shell evolving in the slow-cooling regime with the cooling break above the X-rays \citep{2000ApJ...545..807K}, we would expect the spectral index to be $(p-1)/2$ or $0.65\pm0.13$ for $p=2.32\pm 0.30$; this value of $p$ is explained in more detail in next section.

Figure~\ref{fig:sed} also shows the \emph{Xspec} model fit 
which takes into account the effects of reddening and absorption by the dust \citep{1996ASPC..101...17A}. We use a reddening of E(B-V)=0.02 \citep{2018GCN.22549....1M},
the redshift of $z=2.25$ \citep{2018GCN.22535....1H}, and the column density of $1.77\times10^{20}$cm$^{-2}$ \citep{2018GCN.22540....1P}. We obtained a reduced $\chi^2=0.86$ with 490 degrees of freedom. The $\beta$ derived from the photon index with Xspec is $\beta=0.48\pm 0.01$ which is consistent with the $\beta$ obtained from the simple fit.

\subsection{Temporal Analysis}


The prompt emission from the GRB detected by {\itshape Swift}/BAT lasted until about $T + 120 $~s \citep{22532}. The earliest data  from {\itshape Swift}/XRT started at $T + 61$~s. Our optical observations from TAROT and RATIR began at $T + 26.10$ and $T + 3674$~s, respectively, during prompt emission. Therefore, we focus our analysis on the end of prompt emission and the early afterglow of GRB 180325A.

\subsubsection{Fitting the pulses}
\label{subsec:fitpulse}

\begin{figure}
\centering
  \includegraphics[width=0.45\textwidth]{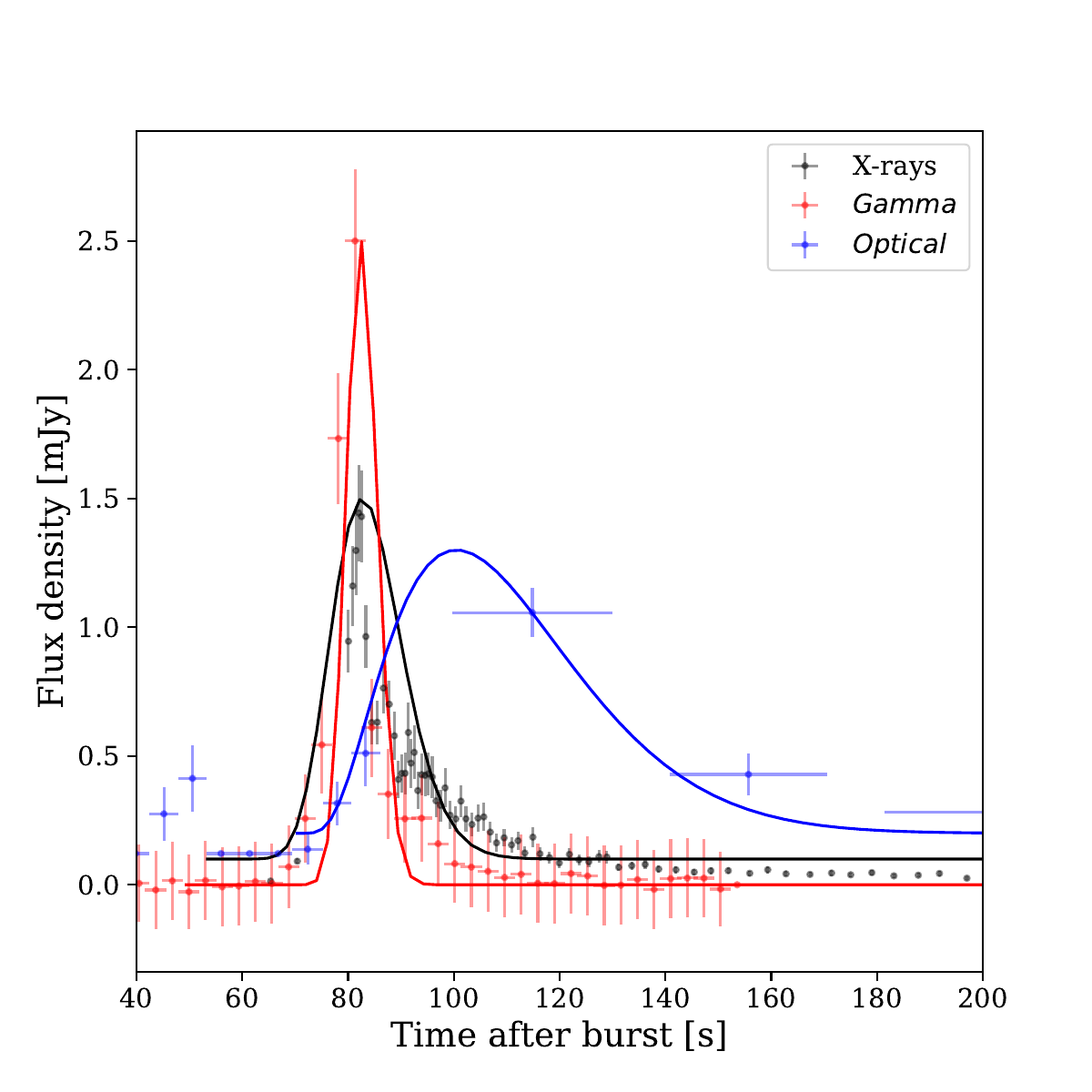}
 \caption{Light curves focused in the end of the prompt emission of GRB 180325A. Data are from {\itshape Swift}, {\itshape Swift}/XRT, and TAROT . Fits of observations were made using the model described by \cite{2005ApJ...627..324N}.}
 \label{fig:early}
 \end{figure}

Figure~\ref{fig:early} shows the early emission in gamma-rays, X-rays, and in the optical for $t<200$~s. 
We fitted these light curves using a model which reproduce the asymmetric shape of the pulses, with a sharper rising phase and a shallower decaying phase. This model, proposed by \cite{2005ApJ...627..324N}, includes an asymmetric exponential rise and exponential-decay profile, and is given by
\begin{equation}
    F(t) = A\; e^{-\tau_1/t - t/\tau_2}\;.
\end{equation}

We notice that the lack of observational points at $T + 70 < t < T + 100$~s makes hard to predict the exact slope of the rising part in the X-rays.

\begin{figure}
\centering
 \includegraphics[width=0.45\textwidth]{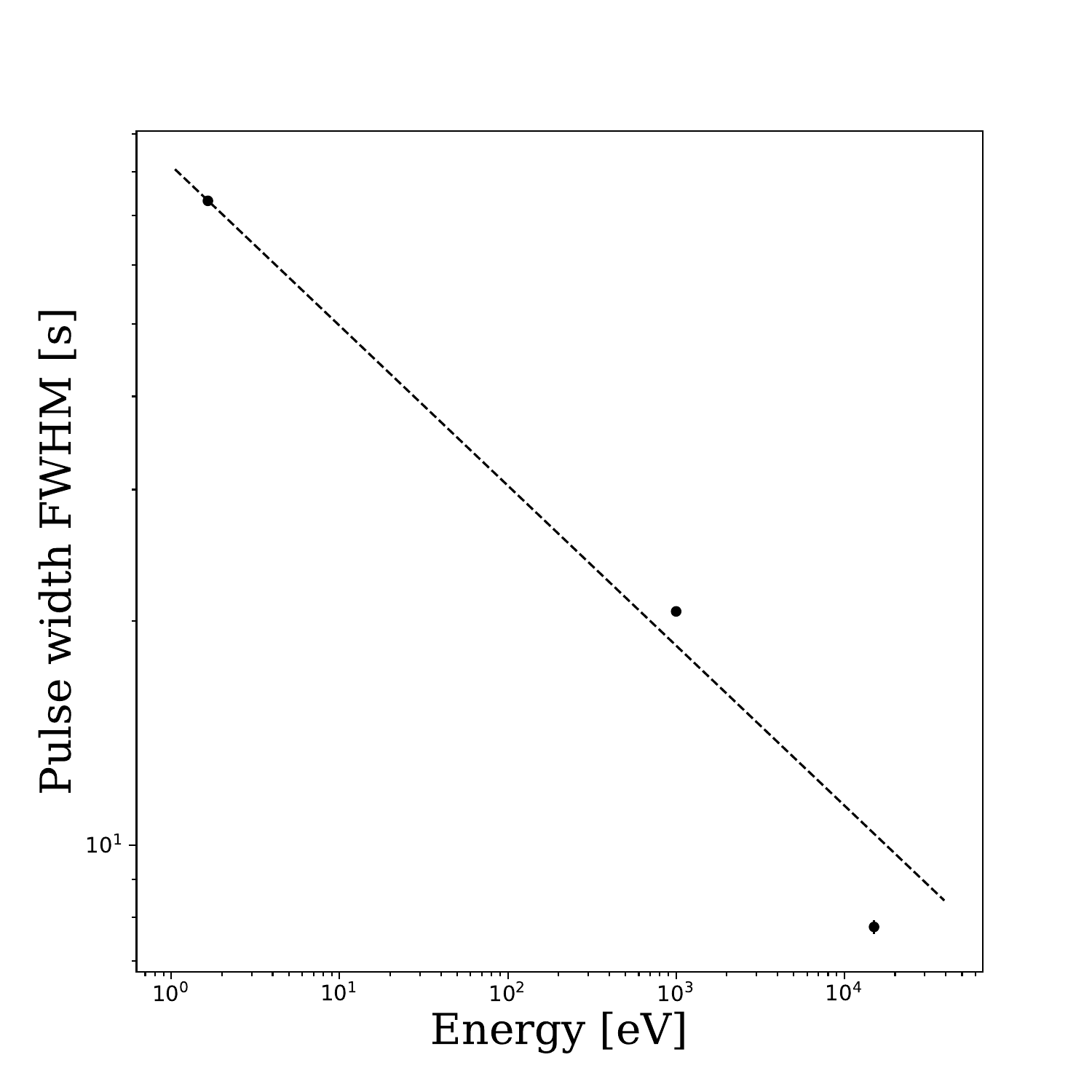}
 \caption{Correlation between the pulse temporal width (measured with the FWHM ) $\omega(E)$ as a function of energy $E$. We found $\omega(E)\propto E^{-0.22\pm 0.03}$}
 \label{fig:fitfit}
\end{figure}

Although, as discussed in section \ref{sec:spike}, the optical emission is likely coming from a reverse shock while the X- and gamma-ray emission originates from internal shock, we fitted the temporal width $w$ for the pulse in the optical, X-rays, and gamma-rays bands using the FWHM. The results are shown in Figure \ref{fig:fitfit}. Then, we fitted the relationship $w(E)\propto E^{-\alpha}$ and found $\alpha=0.22\pm 0.03$ in agreement with $\alpha=0.3$ to $0.4$ found by \cite{2005ApJ...627..324N}  for pulses observed during the prompt emission (between X-rays and gamma rays), and reported by \cite{2010MNRAS.406.2149M} for X-ray flares. If all three pulses have a common origin, then our optical observations would extend this empirical relation down by about three orders of magnitude in frequency. 

Figure~\ref{fig:early} also shows that at $\sim 50$ s there is an earlier optical pulse lasting about 10 s and dimmer by a factor of $\sim 2$ with respect to the late optical peak.   

\subsection{X-rays and Optical afterglow}

For the analysis of the afterglow, we use mainly TAROT and RATIR $r$-band data. The RATIR $r$ and $i$ filters have effective wavelengths of 618 and 760~nm respectively.
We observe very similar 
behaviour between the data for the RATIR $r$ and $i$ filters in Figures~\ref{fig:observations}. We complement our data set with data of the {\itshape Swift}/XRT instrument. 
Figure~\ref{fig:observations} shows the optical and X-ray light curves for GRB 180325A .

We fit the light curves with segments of temporal power-laws $F_\nu \propto t^{-\alpha}$, in which $F_\nu$ is the flux density, $t$ is the time since the BAT trigger, and $\alpha$ is the temporal index. Fits are summarised in Table~\ref{tab:fit}.
We can summarise the main stages as:

\begin{itemize}
\item The light curve for $t < 200$~s. For $ t < 200$~s, we see bright spikes in gamma rays, X-rays, and the optical with a peak at about $T + 100$~s (Figure~\ref{fig:early}). This emission started during the prompt emission phase. Although there is not a completely accepted model to explain these spikes; empirically, we fit two broken power-law segments to the X-ray light curve in order to obtain parameters to describe its behavior. We chose this band because it is the best sampled.  The rise has a temporal index of $\alpha_\mathrm{X,rise}=-7.11$ for $70 < t < 100$~s and the decay has a temporal index of $\alpha_\mathrm{X,decay}=5.42\pm0.26$ for $100 < t < 200$~s. Qualitatively, we note the broadening of the spike as a function of the wavelength.
These phenomena were previously discussed by \cite{2010MNRAS.406.2149M} for X-ray and gamma-ray flares,  with the result that high-energy flare profiles rise faster and decay faster. 
This spike is discussed in more detail in \S~\ref{sec:spike}.

\item The optical curve for $t > 7000$~s. We observed a power-law segment described by $F\propto t^{-\alpha_{f,r/i}}$ with $\alpha_{f,r}=1.48\pm0.18$ for $r$ and  $\alpha_{f,i}=1.49\pm0.23$ for $i$. These temporal indexes can be explained under the assumption of a stellar wind medium where the normal decay segment is described by $F\propto t^{-\alpha_{f,r/i}}$ with $\alpha_{f,r/i}=(1-3p)/4$. Taking $p=2.32\pm 0.30$ (see \S~\ref{sec:spectrum}), the expected value is $\alpha_{f,r/i}=1.52\pm 0.07$ which is in a good agreement with our observed $\alpha_{f,r/i}$.

\item The X-ray light curve for $t > 7000$~s. We observed a power-law segment described by $F\propto t^{-\alpha_{f,X}}$ with $\alpha_{f,X}=1.84\pm0.09$ for the {\itshape Swift}/XRT data. We interpret this as the post jet-break decay phase. In this interval, the optical and X-rays light curves decrease steeper than in the plateau phase. 
We assumed that a non-thermal population of electrons is accelerated by shocks, with a distribution $N(\gamma)\propto \gamma^{-p}$. We analysed this epoch in order to provide a restriction on the power-law index $p$. We then assumed continuity in this population to explain the optical emission during the earlier phase.

Furthermore, from the SED $T+10000$~s, the spectrum can be fitted with a simple power-law with a spectral index of $\beta=0.45\pm0.01$. 
We suggest that this optical emission arises above the cooling break ($\nu_c<\nu$) in a slow-cooling scenario \citep{1998ApJ...497L..17S}, and so expect $F_\nu\propto\nu^{(1-p/2)}t^{(1-3p)/4}$. Our observed value of $\alpha_\mathrm{f,r}=1.49\pm0.22$ implies $p=2.32\pm 0.30$ and subsequently $\beta=0.66 \pm 0.25$. The value of $p$ is consistent with the range of values typically for GRB afterglows and the predicted value of $\beta$ is consistent with our observation of $\beta=0.45\pm0.01$.

\item For $300 < t < 7000$~s, the optical light curve exhibits a plateau with a temporal index of $\alpha_{0,\rm plateau}=0.46\pm 0.01$. The plateau phase is usually interpreted as evidence of late time central-engine activity \citep[see e.g.][]{2015PhR...561....1K,2019ApJ...872..118B,2019ApJ...887..254B}.

\end{itemize}

\section{The origin of the spike}
\label{sec:spike}

\begin{figure}
\centering
 \includegraphics[width=0.45\textwidth]{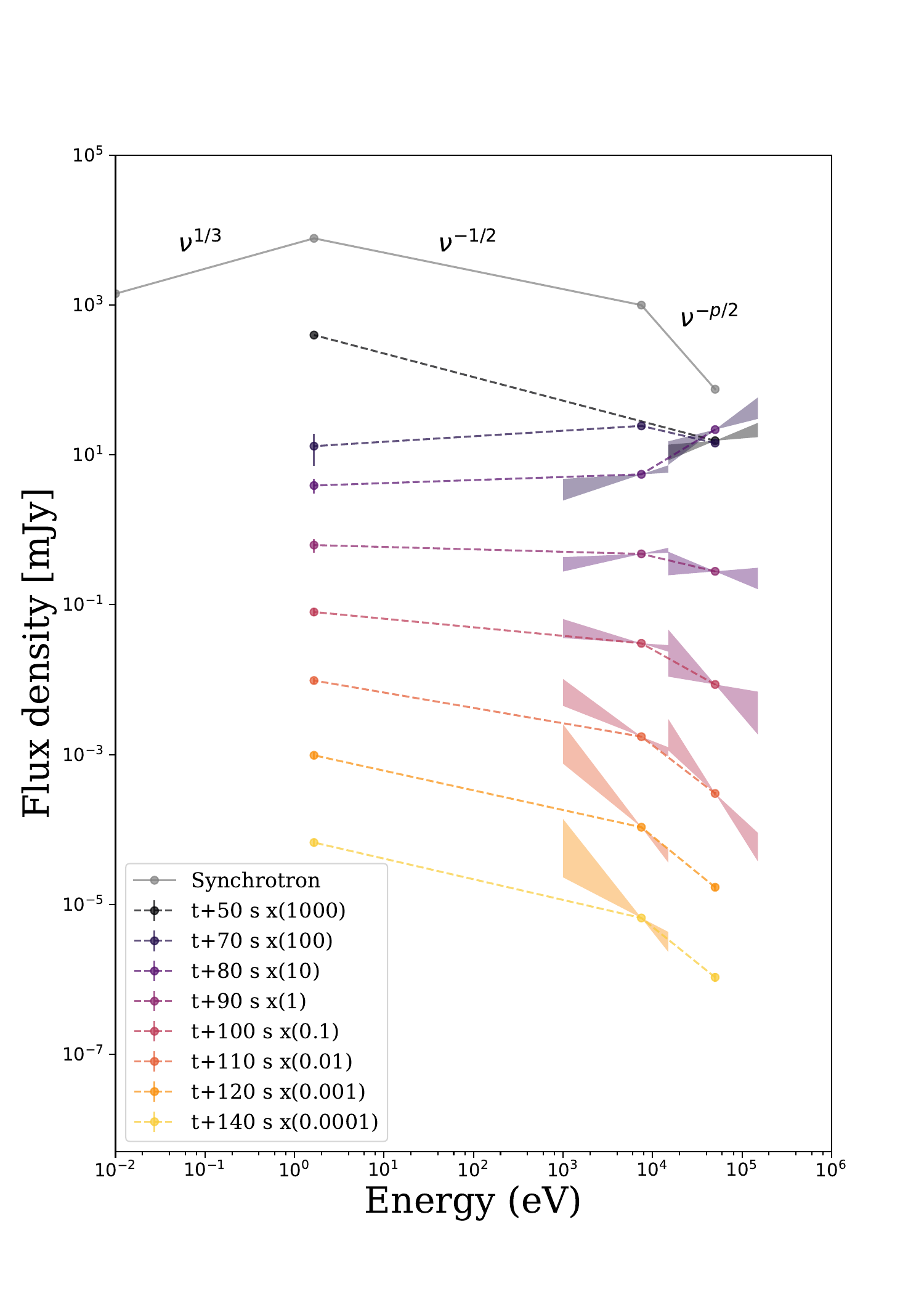}
 \caption{Time evolution of the early time spectrum. The butterfly-scheme is produced using the photon index information retrieved from \itshape{Swift} web page
 }
 \label{fig:spectrum}
\end{figure}

\subsection{Do the optical, X-ray, and gamma-ray spikes have a common origin?}

As we see in Figure~\ref{fig:early}, the sharp peaks become wider at lower frequencies and show a spectral lag, with the high energy light curve rising/dropping on a shorter timescale with respect to the low energy counterpart.
This effect has been observed previously in X-ray/$\gamma$ frequencies in X-ray flares \citep[e.g.,][]{chincarini10,2010MNRAS.406.2149M} and in gamma-ray pulses during the prompt emission \citep[e.g.,][]{1995ApJ...448L.101F, 1996ApJ...459..393N}. 
If the spikes at different frequencies share a common origin, our observations here would extend the broadening relation by three orders of magnitude.

Nevertheless, as shown in Figure \ref{fig:spectrum}, when the optical emission is included the spectrum is concave at $80\;$s after the trigger (near the peak of the light curve in X-rays and gamma-rays), which is not consistent with standard synchrotron radiation. This suggests that the optical spike and the high energy spikes have different origins.
Furthermore, we show below that there is no simple model that can explain the extension of the spectral lag and broadening relations to the optical.
In order to explain the spike we consider the following scenarios.

{\bf High-latitude emission:} 
Spectral lag and broadening have been explained by considering high-latitude emission \citep[e.g.,][]{ ioka01, norris02, dermer04, shen05, genet09, shenoy13}. 
A salient feature of high-latitude emission \citep[e.g.,][]{kumar00,genet09,uhm16,2019ApJ...871..123F,2020ApJ...896...25F} is that
the low-energy flux (at energies well below the $E_p$ of the emission along the line of sight) is much lower than the high-energy flux, as it comes from a region of the shock wave in which the flow is moving slightly off-axis with respect to the observer. Observations of GRB 180325A show that the peak in the flux density, $F_{\nu,p}$, is at a remarkably similar amplitude over four orders of magnitude in frequency. This excludes high latitude emission as a possible explanation.

{\bf Internal shocks:} 
We assume here that the emission is due to synchrotron radiation in the fast-cooling regime, with $F_\nu\propto \nu^{-\alpha}$ for $\nu_c<\nu<\nu_m$, and $F_\nu\propto \nu^{-p/2}$ for $\nu>\nu_m$. 
As discussed by, e.g.,
\citet{daigne11} and \citet{bosnjak14}, there are three possibilities: 1) Synchrotron emission in the fast cooling regime, with $\alpha=1/2$. 2) Fast cooling with $\epsilon_B\ll\epsilon_e$  in the Klein-Nishina (KN) regime\footnote{Here, $\epsilon_B$ and $\epsilon_e$ are the fractions of the post-shock internal energy going into the magnetic field and into the energy of the non-thermal relativistic electrons, respectively.}, where less energetic electrons suffer less KN suppression and radiate more of their energy into the SSC component (that we do not see) and less of their energy into the synchrotron component (that we do see), resulting in a lower flux at lower frequencies compared to case 1, i.e., a lower $\alpha$: $0\lesssim\alpha\leq1/2$. 3) Marginally fast cooling, $\nu_c\approx\nu_m$, with $\alpha=1/3$ corresponding to a different power-law segment of the synchrotron spectrum.
In all these cases, the flux typically shows some spectral lag and broadening in the X-rays and gamma-rays range. 
However, it is unclear if this behaviour can be extended to the optical range, as a drop in the peak energy (due, e.g., to a drop in the microphysical parameters $\epsilon_e$, $\epsilon_B$, or to the shock Lorentz factor, or adiabatic cooling) should lead to a dramatic drop in the flux density. Moreover, the optical emission may also be self-absorbed, leading to a further suppression in its flux\footnote{ Self-absorption should be less important in reverse shocks, as their density is much lower.}.

{\bf An accelerating jet coupled with a decaying magnetic field:}
\citet{uhm16} interpreted the spectral lag and broadening as due to a decaying magnetic field in an accelerating emitting region (and with a curved emitted photon spectrum in the proper frame). They point out that this is consistent with a Poynting-flux dominated jet accelerated far away from the source \citep[e.g.][]{drenkhahn02,granot11}. As the synchrotron flux drops with the magnetic field, an increase in the shock velocity keeps the low-energy flux density at a level comparable with the high-energy component. 

Following \citet{uhm16}, the peak energy of the observed synchrotron spectrum is $E_p\propto \Gamma B$ (where $\Gamma$ is the Lorentz factor of the radiating material), and the peak flux scales as $F_{\nu,p} \propto N \Gamma B$, where $N$ ($\propto r/\Gamma^2$) is the number of electrons assumed to be constantly injected in the emitting region. Thus, considering a magnetic field in the comoving frame dropping as $B\propto r^{-b}$, and a Lorentz factor increasing as $\Gamma\propto r^s$, \citet{uhm16} got $E_p\propto r^{s-b} \propto t_{\rm obs}^{(s-b)/(1-2s)}$ (where, to derive the last proportionality we employed the relation $t_{\rm obs}\propto t/\Gamma^2 \propto r/\Gamma^2 \propto r^{1-2s}$) and $F_{\nu,p} \propto t_{\rm obs}^{(1-b)/(1-2s)}$. Finally, we obtain
$F_{\nu,p} \propto E_{p,\rm obs}^{(1-b)/(s-b)}$. As the peak energy should drop by about five orders of magnitude in the emitting region (going from $\sim200\,$keV to $\sim2\,$eV), the exponent $(1-b)/(s-b)\approx 0$ to get a nearly constant flux density as was observed. This implies that $b\approx 1$. For such a drop in the comoving magnetic field, $E_p\propto \Gamma B\propto r^{s-1}$ or $E_p\propto r^{-2/3}\propto\Gamma^{-2}$ for $s=1/3$ that is expected in Poynting-flux dominated models, which implies that the emitting region should extend over $\sim7.5$ decades (e.g., from 10$^{10.5}\;$cm to 10$^{18}\;$cm), while $\Gamma$ increases by a factor of $\sim10^{2.5}$. Given that even at the initial radius $\Gamma\gtrsim10^2$ is typically required because of compactness arguments, this would require an unreasonably large $\Gamma\gtrsim10^{4.5}$ near the peak time of the optical emission. 

In conclusion, while the possibility that the optical emission shares a common origin with X- and gamma-rays cannot be completely ruled out, it is hard to explain with synchrotron emission. There are further difficulties in explaining
the broadening over such a large frequency range.
Moreover, a common origin does not naturally explain the convex spectrum shown in Fig.~\ref{fig:spectrum}, which suggests a distinct spectral component in the optical.

\subsection{Optical, X-rays and gamma-rays originated from reverse and internal shocks}

Finally, we consider the possibility that the optical emission originates from the external reverse shock\footnote{ Optical flares can also be produced by synchrotron emission from thermal $e^\pm$ pairs
behind the forward shock, as proposed by \cite{2014ApJ...789L..37V} to explain the optical flash in GRB 130427.} that decelerates the ejecta as it sweeps up the external medium, while the X-rays and gamma-rays are from prompt emission (e.g., produced by internal shocks or magnetic reconnection within the ejecta).

A reverse shock component is expected as the expanding relativistic ejecta encounters the circumburst matter (CBM; be it the ISM for short-hard GRBs or the stellar wind of the massive progenitor star for long-soft GRBs). A forward shock is driven into the CBM, a reverse shock is driven upstream into the unshocked part of the ejecta, and the two shocked regions are separated by a contact discontinuity. A reverse shock is predicted to produce a single peak in the light curve, which may appear similar to a flare \citep[see, e.g.,][]{SP99,2007ApJ...655..391K}, as the electrons are heated and cool primarily by synchrotron radiation and inverse-Compton scattering of the synchrotron photons (synchrotron self-Compton; SSC). 
\cite{2007ApJ...655..391K} study the variability and the temporal indexes expected in a X-ray emission created in the reverse shock region. A reverse shock must display a time variability scale of $\Delta t/t\sim 1$  \citep{2007ApJ...655..391K} and varies as $F_\nu\propto t^{5(p-1)/4}$ before the peak and $\propto t^{-(3p+1)/3}$ after the peak. We observe a $\Delta t=130$~s and $t=100$~s in the optical band and therefore $\Delta t/t\sim 1.3$, consistent with a reverse shock component. 

It is useful to compare the temporal indices expected for this model to our observations in X- and gamma-rays. We divide the X-ray sharp peak in two regions to fit them with power-law segments. From $T+70$~s to $T+200$~s with a peak in $T+100$~s, we found temporal indexes of $\alpha_{\rm X, rise}=-7.11\pm2.77$ and $\alpha_{\rm X,decay}=5.42\pm2.54$ for the rise and decay respectively. Using the value of $p$ described in \S~\ref{sec:analysis} of $p=2.32 \pm 0.30$ for the afterglow phase, we expect a $F_\nu\propto t^{5(p-1)/4}=t^{1.65}$ before the peak and $\propto t^{-(3p+1)/3}=t^{-2.65}$ after the peak. The observed values are quite different to those expected theoretically for $\alpha_{\rm X, rise}$ and $\alpha_{\rm X,decay}$. Therefore, we can conclude that a reverse shock (at the jet head) is not a plausible mechanism for the X-ray sharp peak of GRB 180325A, which likely originated from internal shocks.

On the other hand, early optical afterglows often have an external reverse shock (RS) origin \citep{1993ApJ...405..278M}. In addition, a RS  may explain the temporal behaviour of the optical light curve in GRB 180325A. In particular, a peak on a time comparable to and slightly longer than the duration of the prompt GRB emission is expected for a reverse shock that is at least mildly relativistic -- the ``thick shell'' case, and the observed optical light curve may correspond to a mildly relativistic RS \citep[see, e.g.,][]{nakar04}.


\begin{figure}
\centering
\includegraphics[width=0.5\textwidth]{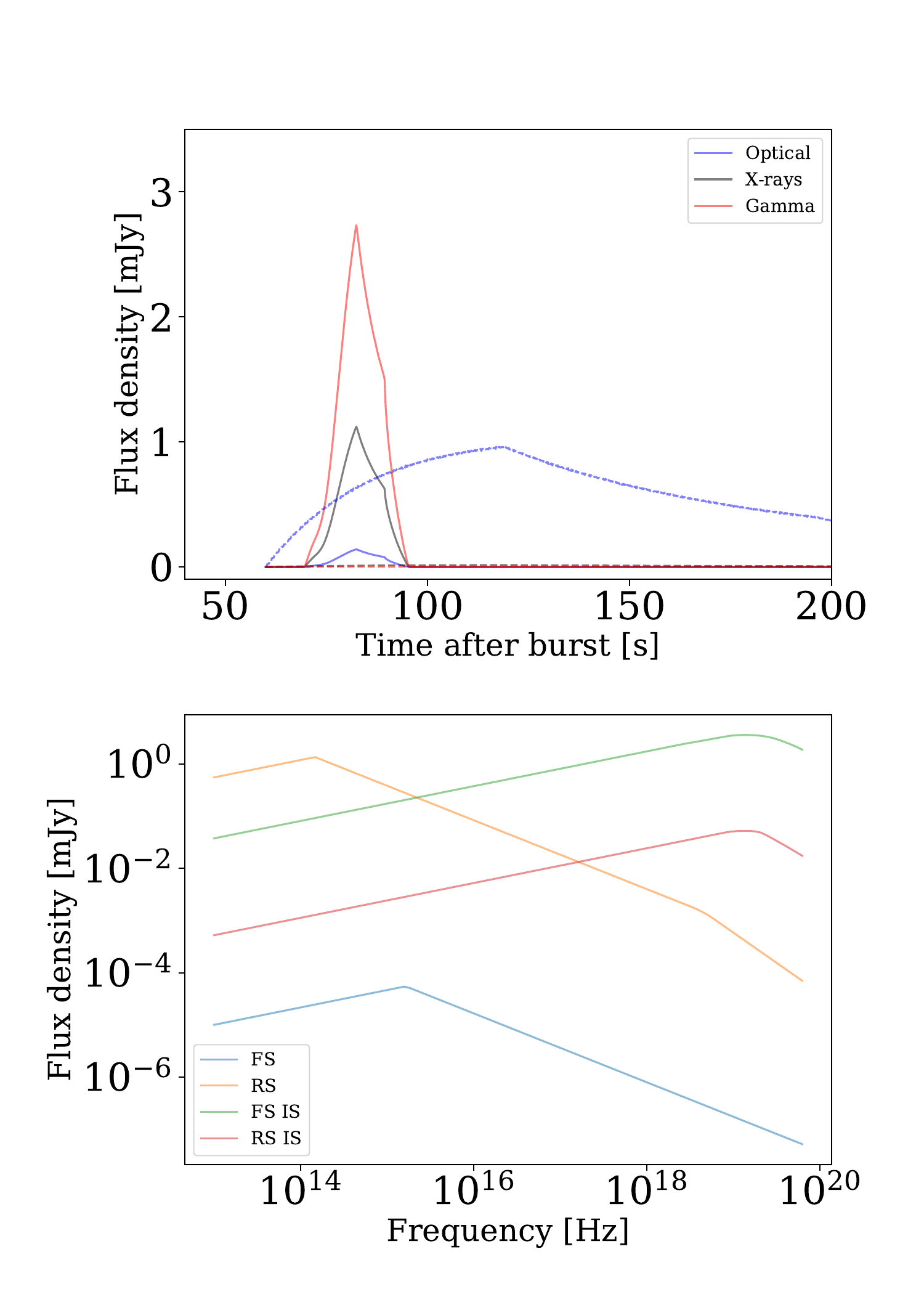}
\caption{
\emph{Upper panel:} Light curves (in three energy bands) emitted from the working surface as result of the interaction between two internal shocks (solid lines) and by the (external) reverse shock (dashed lines). Synchrotron emission is assumed in all the cases. X- and gamma-rays produced by the RS are negligible, and the optical emission produced by the IS is smaller than the corresponding RS component. The observed increase in flux associated with the IS is likely due to the shells not being uniform in density as assumed in our model. \emph{Bottom panel:} Spectra emitted by the reverse and forward shock formed by internal shocks (at $t=82$ s, ``RS IS'' and ``FS IS'' in the figure), and by the reverse and forward shock at the jet head (at $t=120$ s, ``RS'' and ``FS'' in the figure). The RS dominates the emission in both cases.}
\label{fig:model}
\end{figure}

To reproduce the observations, we assume synchrotron radiation and consider the emission of the two components, i.e. internal and external shocks.  Internal shocks result from a non-uniform distribution of the Lorentz factor in the outflow, while by ``external shocks'' we refer to both the forward shock which propagate into the external medium and the reverse shock that sweeps back into the ejecta as a result of the deceleration of the jet material.
We employ the formalism of \cite{2013MNRAS.430.2703C} for the dynamics of internal shocks in relativistic jets with a time-dependent injection velocity and mass-loss rate. This formalism takes into account the loss of 
energy and momentum due to radiation. \cite{2013MNRAS.430.2703C} found an analytic solution for the dynamics of the internal shocks  in the case where the material is ejected from the central engine with 
a step function velocity variation. 

We consider the interaction between two shells with normalised velocities $\beta=v/c$, where $c$ is the speed of light, and Lorentz factors $\gamma$. 
We assume that the first shell is ejected with a slow velocity $\beta_s$ during a time interval $\Delta \tau_s$, with a mass injection rate $\dot{m}_s$. After a time interval 
$\Delta \tau_0$ a second shell, with a fast velocity $\beta_f$ {\bf $> \beta_s$} and mass injection rate $\dot{m}_f$, is injected during a time interval $\Delta \tau_f$.  The two shells collide after a time  $t_{\rm coll} =\beta_f \Delta \tau_0/(\beta_f - \beta_s) \approx 2 \gamma_s^2 \Delta \tau_0/(1-\gamma_s^2/\gamma_f^2)$ (measured from the time when the ejection of the first shell ended) at a distance $x_{\rm coll} = c \beta_s t_{\rm coll}$. 
At this moment, a two shock structure (the \emph{working surface}, WS hereafter), is formed at $x_{\rm coll}$. 

In addition, by assuming $\beta_s\ll 1$, the same formalism allows us to study the propagation of the jet head (with the WS made in this case by the forward and reverse shocks).  To apply this formalism, one has to substitute in their equations the injection time of the fast material $\tau_f$ by $\tau_f - \Delta \tau_0$.  This change does not affect the shell dynamics and is equivalent to a reference frame where  $x_{\rm coll} = t_{\rm coll} = 0$,  and the gas particles are injected at $x_{\rm coll} =0$ with a velocity distribution shown in Figure \ref{fig7}. In  Appendix~\ref{app:step} we summarise their solution.

The interaction between the internal shocks moving into the jet channel leads to the formation of a forward shock, which propagates into the slow medium incorporating this material into the WS, and a reverse shock  decelerates and incorporates the fast moving material (in the WS frame, the reverse shock crosses the fast moving material). This double shock structure is similar to the one formed at the external shocks, although in internal shocks both the fast and slow material move with relativistic speeds, while in the external shocks the slow moving material is moving at a much smaller speed (being the wind of the progenitor star).

Light curves computed by employing the semi-analytic model described above (see Appendix~\ref{app:sync}) are shown in Figure~\ref{fig:model}. We take $\xi_e=0.1$ (external shocks) and  $\xi_e=10^{-3}$ (internal shocks), being $\xi_e$ the fraction of electrons accelerated at the shock front.
The light curves are produced by considering standard synchrotron theory with $\epsilon_e=0.1$, $\epsilon_B=10^{-2}$ and $p=2.32$ for both components, RS and forwards shock (FS). 

The high-energy component is due to internal shocks, with  a Lorentz factors of $\gamma_f=200$ and $\gamma_s=30$ for the fast and slow shell respectively.  We consider an isotropic energy $E=2.8\times 10^{53}$ erg for both the slow and fast shell.
The shells are injected during a time $\Delta \tau_f = \Delta \tau_s=$ 2 s, with mass-loss rates $\dot{m}_s=\dot{m}_f$ (sub indices \emph{s} and \emph{f} refer to slow and fast components respectively).
The shells collide at $R\sim 10^{15}$ cm and form a working surface which moves with a Lorentz factor $\sim 100$.
The energy of the fast and slow shells is
\begin{equation}
E_{(s,f)} = \dot m_{(s,f)} \Delta \tau_{(s,f)} \gamma_{(s,f)} c^2\;.
\end{equation}
Given the values of $\Delta \tau_{(s,f)}$ and $\gamma_{(s,f)}$ specified above, this equation gives the explicit values of $\dot m_{(s,f)}$.

On the other hand, the low-energy component is due to the reverse shock. The slow-moving material has $\gamma_s = 1.0017$ (corresponding to a typical Wolf-Rayet wind velocity of $v_w=10^8$ cm s$^{-1}$), $\gamma_f=30$, $\dot{m}_s=10^{-6}$ M$_\sun$ yr$^{-1}$, $\Delta t_s=\infty$, $\Delta t_f=1$, $E_f=4\times 10^{52}$ erg, and $\dot{m}_f=E_f/(\Delta \tau_f \gamma_f c^2)$.   

In the slow cooling regime, the synchrotron emission depends strongly on the value of $\nu_m$ (i.e. the peak frequency). Then, to explain the observations with our model we need to choose fitting parameters such that $\nu_m$ peaks close to optical frequencies in the reverse shock and to gamma frequencies in internal shocks (see Figure~\ref{fig:model}, bottom panel). That implies that: 1) the fraction of accelerated electrons $\xi_e$ in the RS is much larger then in IS (as $\nu_m\propto 1/\xi_e^2$); or 2) the jet luminosity/ejecta energy is much larger in the IS than in the RS.
In the first case, different values of $\xi_e$ in the IS and RS are typically considered in the literature. Both IS and the RS propagate in the ejecta, but shocks in IS are relativistic while the RS is only mildly relativistic. Then, the acceleration process is much more efficient in RS than in IS.

While the optical light curve shown in Figure~\ref{fig:model} accounts for the main optical peak at $\sim120\;$s, it does not account for the earlier optical peak at $\sim50\;$s and the associated temporal variability that is reflected in these two peaks. This variability may potentially arise from the density structure within the ejecta that is encountered by the reverse shock, which is caused by the collision and merger of the original shells that produced the internal shocks responsible for the X-ray and gamma-ray prompt emission. This is in contrast to the model described above for the reverse shock, which assumes a perfectly uniform ejecta shell and therefore leads to a single peak in the light curve.  An additional contribution to the time variability of the optical emission from the reverse shock may arise from a variation of the magnetization within the ejecta, where strongly magnetized portions suppress the reverse shock and its emission while mildly magnetized portions lead to a strong reverse shock with prominent emission.

\section{Summary and discussion}
\label{sec:discussion}

We have presented optical photometry of the very early afterglow of GRB 180325A with the TAROT instrument from $T + 26$~s. We complement these data with RATIR observations up to  30 hours after the burst. 

We compare the optical light curve 
with gamma- and X-ray light curves from {\itshape Swift}/BAT and {\itshape Swift}/XRT respectively. We see an early gamma/X-ray/optical spike from $T + 70$~s, with a peak at about $T + 100$~s, and lasting until about $T + 200$~s. The peak at different times and with not the same duration, turns to GRB 180325A in a very special case of study. 


By discussing possible scenarios for its origin we show that the  X-ray/gamma-ray spike is likely have been generated by an internal shock while the optical spike is likely to have been generated by an external reverse shock. Interestingly, optical observations show an earlier flare, lasting about 10 s, which is also not associated with the high energy emission and is possibly also the result of a reverse shock. 
Observations of fluctuations in the optical reverse shock emission is atypical and might be the result of  density/velocity fluctuations in the material crossing the reverse shock.

\section*{acknowledgments}

We thank the referee for useful comments that helped to improve this article. We thank Fr\'ederic Daigne for useful discussions.
We thank the staff of the Observatorio Astron\'omico Nacional on Sierra San Pedro M\'artir.
RATIR is a collaboration between the University of California, the Universidad Nacional Auton\'oma de M\'exico, NASA Goddard Space Flight Center, and Arizona State University, benefiting from the loan of an H2RG detector and hardware and software support from Teledyne Scientific and Imaging. RATIR, the automation of the Harold L. Johnson Telescope of the Observatorio Astron\'omico Nacional on Sierra San Pedro M\'artir, and the operation of both are funded through NASA grants NNX09AH71G, NNX09AT02G, NNX10AI27G, and NNX12AE66G, CONACyT grants INFR-2009-01-122785 and CB-2008-101958, UNAM PAPIIT grants IG100414, IA102917, UC MEXUS-CONACyT grant CN 09-283, and the Instituto de Astronom{\'\i}a of the Universidad Nacional Auton\'oma de M\'exico. We acknowledge the vital contributions of Neil Gehrels and Leonid Georgiev to the early development of RATIR. 
TAROT has been built with the support of the Institut National des Sciences de l'Univers, CNRS, France. TAROT is funded by the CNES and thanks the help of the technical staff of the Observatoire de Haute Provence, OSU-Pytheas.
This work made use of data supplied by the UK Swift Science Data Centre at the University of Leicester.
RLB aknowledges support from the DGAPA-UNAM postdoctoral fellowship.
We aknowledge support from the UNAM-PAPIIT grant AG100820 and AG100317. JC, SL, NF and RFG acknowledge support from PAPIIT/UNAM IG100218, IN101418, IA102019,   IN107120. JG aknowledges support from the ISF-NSFC joint research program (grant No. 3296/19). ATA thanks the Czech Science Foundation under the grant GA\v{C}R 20-19854S.

\begin{deluxetable*}{lccrrr}
\tabletypesize{\scriptsize}
\tablewidth{0pt}
\tablecaption{Fitting parameters of early data GRB 180325A \label{tab:parameters}}
\tablehead{\colhead{Regime}&\colhead{$A$}&\colhead{$\tau_1$}&\colhead{$\tau_2$}&\colhead{$\tau_s$}&\colhead{$FWHM$ [s]}}
\startdata
Gamma&2.5&5000.0&2&50&7.77\\
X-ray&1.4&1000&5&55&20.59\\
Optical&0.8&100.0&100.0&80.0&73.19\\
\enddata
\end{deluxetable*}

\begin{deluxetable*}{lcr}
\tabletypesize{\scriptsize}
\tablewidth{0pt}
\tablecaption{Fitting parameters of GRB 180325A \label{tab:fit}}
\tablehead{\colhead{}&\colhead{Parameter}&{Value}}
\startdata
\textbf{General}&&\\[2ex]
\hline
\hline\\
Electron index&p&$2.36\pm0.30$\\[2ex]
\hline
\hline\\
\textbf{Plateau}& ($300<t<7000$ s)&\\[2ex]
\hline
\hline\\
Optical&$\alpha_\mathrm{p,r}$&$0.46\pm0.01$\\[2ex]
\hline
\hline\\
\textbf{Forward shock}& ($t>7000$ s)&\\[2ex]
\hline
\hline\\
Optical&$\alpha_\mathrm{f,r}$&$1.48\pm0.38$\\
&$\alpha_\mathrm{f,i}$&$1.49\pm 0.01$\\
$X$-rays&$\alpha_\mathrm{f,X}$&$2.32\pm 0.20$\\
&$\beta$&$0.69\pm0.01$\\[2ex]
\enddata
\end{deluxetable*}



\newpage
\renewcommand{\thesection}{\Alph{section}}
\setcounter{section}{0}

\section{Calculation of light curve and spectrum}

Given the dynamical evolution of the forward and reverse shocks, we compute the multi-wavelength emission by assuming that it is produced by synchrotron radiation. We discuss here the dynamical evolution of the working surface, and how it is used to compute the emitted radiation.

\subsection{Step function velocity variation} 
\label{app:step}

\begin{figure}
\centering
\includegraphics[angle=0,width=0.45\textwidth]{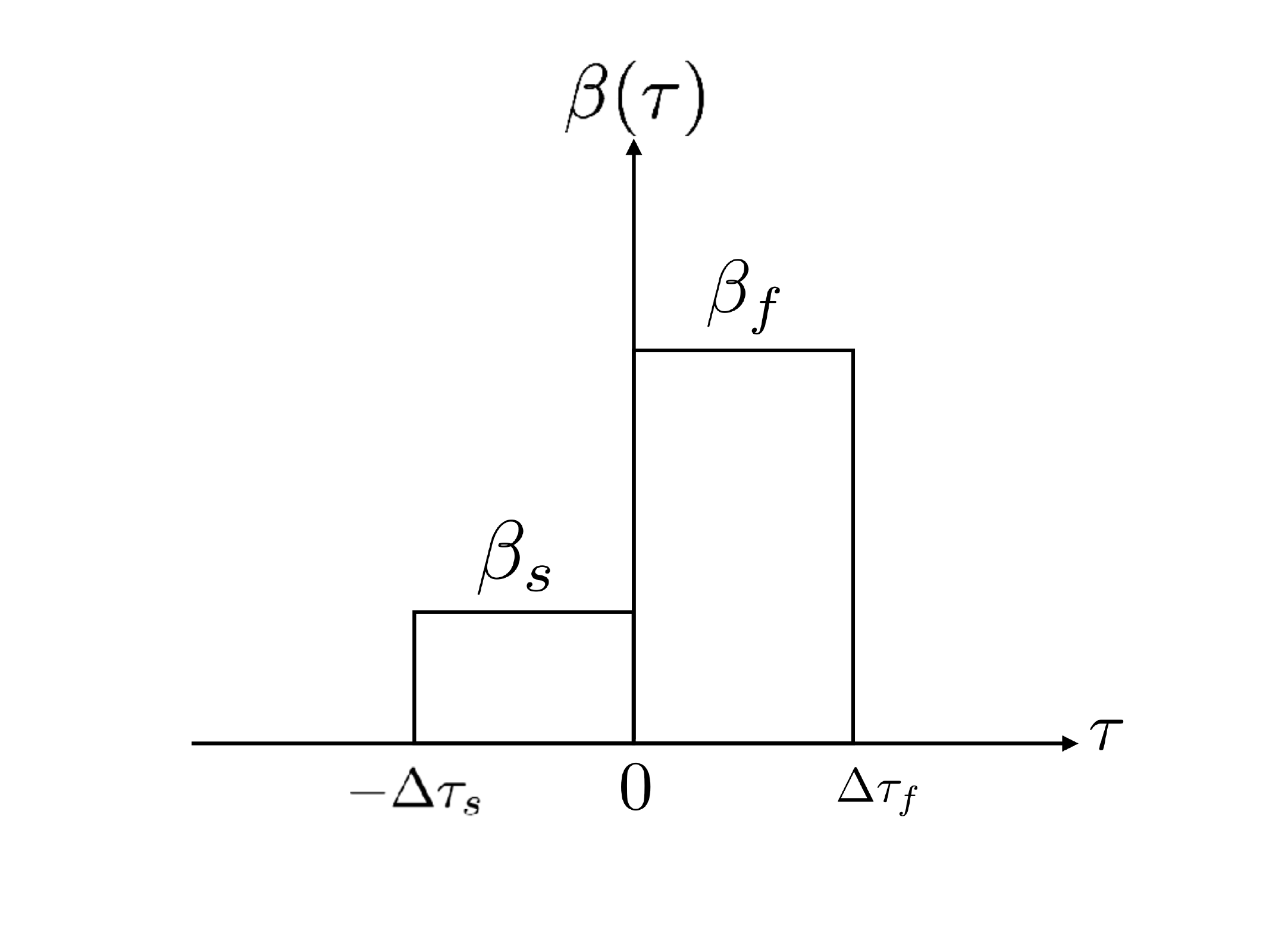} 
\caption{Injection velocity $\beta(\tau)$ as a function of time $\tau$ at the position of the shell collision. The time intervals 
$\Delta \tau_s$ and $\Delta \tau_f$ represent the duration of the injection of the fast and slow material.}
\label{fig7}
\end{figure}

Following  \cite{2013MNRAS.430.2703C},  we consider a free-streaming flow injected at $x=0$ with velocity $\beta$ and mass-loss rate $\dot m$. We assume a velocity variation (see Figure \ref{fig7}) with a slow flow injected in the interval $[-\Delta \tau_s, 0]$ and
 a fast flow injected in the interval $[0, \Delta \tau_f]$.
 A working surface (WS) is formed at the position $x_{\rm ws}=0$ at a time $t=0$.


Initially, the WS velocity is constant, and corresponds to a Lorentz factor
\begin{equation}
\gamma_{ws0}=
\frac{\lambda+1}{ 
\sqrt{ \left( {\lambda/ \gamma_f} \right)^2  + 2 \lambda\left(1-\beta_s\beta_f\right) +
1/\gamma_s^2  } } .
\label{eq:gamma0}
\end{equation}
In this equation one defines
\begin{equation}
\lambda = \sqrt{\frac{b r}{a}},
\label{eq:lambda}
\end{equation}
where the velocity ratio, the mass-loss ratio, and the gamma ratio are given by
\begin{equation}
 a=\beta_f/\beta_s, \quad b=\dot m_f/\dot m_s, 
\quad r= \gamma_f/\gamma_s.
\end{equation}

\begin{figure}
\centering
\includegraphics[width=0.5\textwidth]{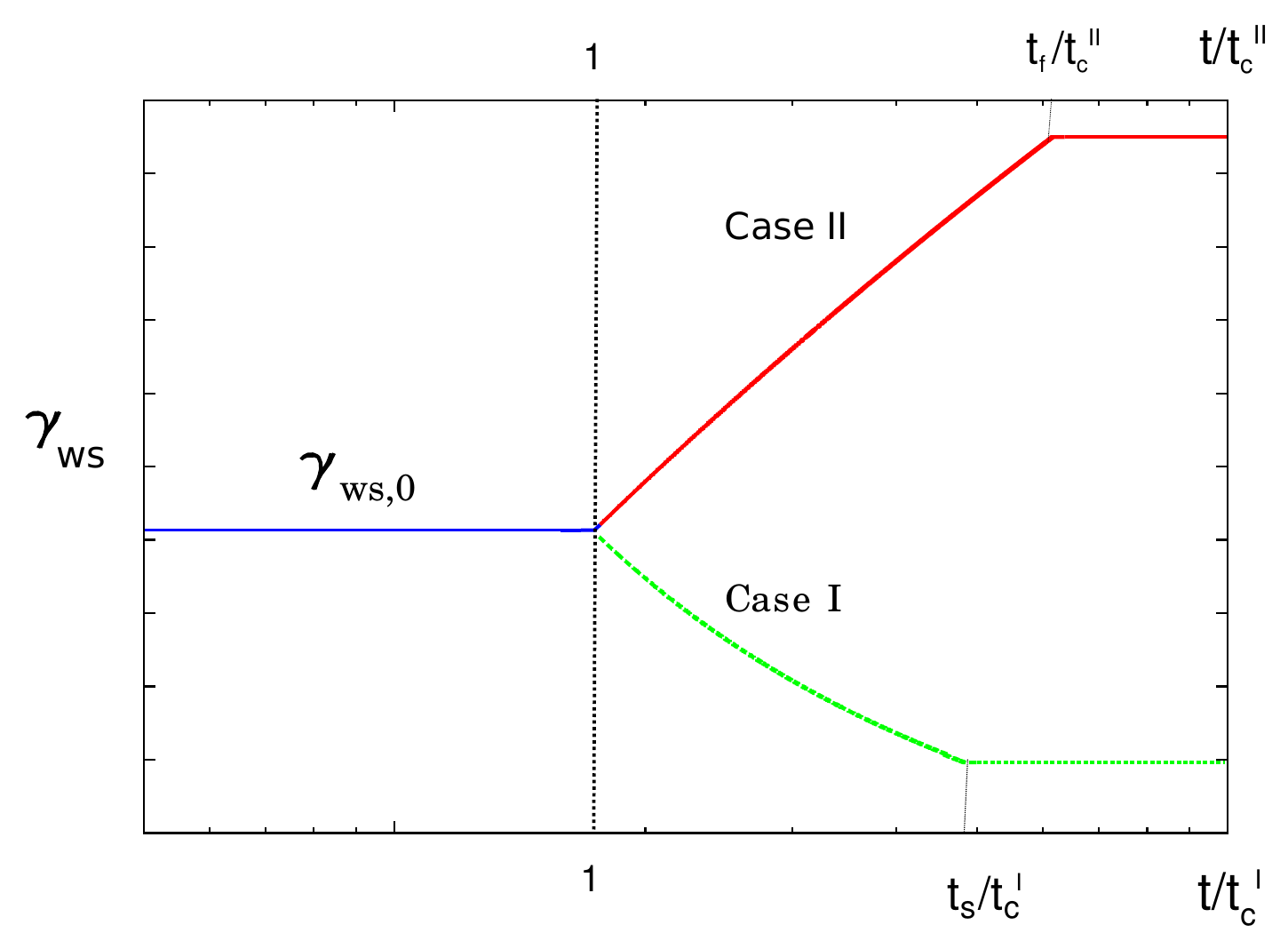}
\caption{Qualitative temporal evolution of the working surface Lorentz factor for Case I (green line) and Case II (red line). The initial constant velocity for the two cases is shown in blue. 
}
\label{fig8}
\end{figure}

The velocity remains constant during a certain time, which depends on which material runs out first.
In case I (see Figure \ref{fig8}), the constant velocity phase ends when the fast material is completely incorporated into  the WS, at a time $ t_{\rm c}^{I}$  given by 
\begin{equation}
 t_{\rm c}^{I} = \frac{a \left( \lambda + 1 \right)}{a-1} \Delta \tau_f
 \approx
  \frac{2\gamma_s^2 }{1-\gamma_s^2/\gamma_f^2}  \left( \lambda + 1 \right) \Delta \tau_f .
\label{tcI}
\end{equation}
Afterwards, the WS decelerates such that the 
Lorentz factor  $\gamma_{\rm ws}^I(t)$ varies with time as
\newcommand{\T}{\mathcal{T}}
\begin{equation}
\gamma_{\rm ws}^I= \gamma_s \left( \frac{\T^2 + 4 \beta_s \T + 4}{\T^2 - 4 } \right),
\label{eq:gammaI}
\end{equation}
where $\T(\tilde t)$ is a function of time given by the transcendental equation (23) in \cite{2013MNRAS.430.2703C},
where $\tilde t = t / \Delta \tau_f$.\footnote{Note that in \cite{2013MNRAS.430.2703C} subscript 1/2 refers to the slow/fast material.}
 The deceleration phase ends at a time $t_s$, when the WS completely engulfs the slow moving material. For a 
 free-streaming flow, $t_s$ is given by the condition
\begin{equation}
 v_s \;(t_s + \Delta \tau_s) =
  R_{\rm ws}(t_s).
\end{equation}

 In case II, the constant velocity phase ends
 when the slow material is completely incorporated into  the WS, at a time $ t_{\rm c}^{II}$
 given by 
  \begin{equation}
 t_{\rm c}^{II} = \frac{  \lambda + 1}{\lambda (a-1)}\Delta \tau_s
 \approx
  \frac{2\gamma_s^2}{1-\gamma_s^2/\gamma_f^2}  \frac{\lambda + 1}{\lambda}  \Delta \tau_s
\label{tcII}
\end{equation}
 The WS then accelerates  such that
the Lorentz factor $\gamma_{\rm ws}^{II}(t)$ is
\begin{equation}
\gamma_{\rm ws}^{II}= \gamma_f \left( \frac{\T^2 - 4 \beta_f \T + 4}{\T^2 - 4 } \right),
\label{eq:gammaII}
\end{equation}
where $\T(\tilde t)$ is given by the transcendental equation (35) in \cite{2013MNRAS.430.2703C},
where  $\tilde t = t / \Delta \tau_s$.
The acceleration phase ends at a time $t_f$, when the WS completely engulfs the fast moving material, i.e. when 
\begin{equation}
 v_{f} \;(t_{f} - \Delta \tau_{f}) = R_{\rm ws} (t_s).
 \label{eq:tfast}
\end{equation}

Finally, the velocity of the WS becomes constant when all the material has been incorporated (see Figure \ref{fig8} for a schematic representation of the evolution of the Lorentz factor of the WS). 

In the case of a fast shell that runs into a static medium  with a density $\rho_s$ and $\beta_s =0$, since the mass-loss-rate is $\dot m \propto \rho \beta$ , the parameter $\lambda$
 can be written as $  \lambda = (\rho_f \gamma_f/\rho_s)^{1/2}$. Then, the phase of constant velocity has
 \begin{equation}
 \gamma_{ws0} = \frac{\lambda+1}{ 
\sqrt{ \left( {\lambda/ \gamma_f} \right)^2  + 2 \lambda +
1 } } .
 \end{equation}
 This phase ends at the time $t_c^I$ when all the fast material has been incorporated to the WS. From then on,
 the WS decelerates continuously with a Lorentz factor given by eq. (\ref{eq:gammaI}) and approaches asymptotically 
 to rest (see case I in Fig. 3 of \citealt{2013MNRAS.430.2703C}).

\subsection{Synchrotron emission}
\label{app:sync}

We solve the Taub relativistic shock-jump conditions and compute the post-shock energy density $e_{\rm ps}$, density $\rho_{\rm ps}$ and velocity $v_{\rm ps}$ as a function of the WS velocity, the pre-shock densities and the velocities of the fast and slow material ejected from the central engine. Thus, the post-shock physical variables will evolve with time as the working surface accelerates or decelerates as discussed in Appendix \ref{app:step}.

To compute the synchrotron radiation, we assume that a fraction $\chi_e$ of the post-shock electrons are accelerated to relativistic speeds, forming a population of electrons $N(\gamma_e) \propto \gamma_e^{-p}$, where $p$ is the power-index of the population of non-thermal electrons accelerated by the shock.   The energy density of the accelerated electrons and of the post-shock magnetic field are taken as a fraction $\epsilon_e$ and $\epsilon_B$ of the post-shock thermal energy density. Then, we use standard synchrotron theory to compute the synchrotron emissivity \citep[see, e.g.,][]{decolle12} as a function of time.

In synthesis, light curve and spectrum are computed as a function of the following parameters: the Lorentz factors $\gamma_f(t)$ and $\gamma_s(t)$ of the fast and slow unshocked material, the mass-loss rates (or the density) $\dot{m}_f$,  $\dot{m}_s$, and the duration of injection of the fast and slow material $\Delta \tau_f$ and $\Delta \tau_s$.  In addition, the synchrotron radiation calculation is parameterized as a function of the microphysical parameters $\epsilon_e$, $\epsilon_B$ and $p$.

\end{document}